\documentstyle[12pt]{article}

\begin{document}

\begin{flushright}
{\footnotesize HEP-TH/yymmdd - IASSNS-HEP-95/45}
\end{flushright}

\vspace*{5cm}

\begin{center}

{\Large \bf QUATERNIONS FOR GUTs}

\vspace*{2cm}

STEFANO DE LEO\\

\vspace*{1cm}

{\it Institute for Advanced Study\\ Princeton, NJ 08540 USA}\\
{\footnotesize and}\\
{\it Universit\`a  di Lecce, Dipartimento di Fisica\\
I.~N.~F.~N. - Lecce, Italy}

\vspace*{2cm}

{\bf Abstract}

\end{center}
We derive an appropriate definition of transpose for quaternionic matrices and
give a new panoramic review of the quaternionic groups. We aim to analyse
possible quaternionic groups for GUTs.

\pagebreak

\section{Introduction}

This paper aims to give a clear and succint classification of
possible quaternionic groups for Grand Unification Theories. We know that the
standard group theory which applies to elementary particle physics is
understood to be complex, nevertheless we must observe that non-supersymmetric
GUTs based on complex groups have run into difficulties. A stimulating
possibility~\cite{adl} is that a
successful unification of the fundamental forces will require a generalization
beyond the complex.

We will discuss in this paper unitary, special unitary,
orthogonal (a new  definition of transpose for quaternionic matrices overcomes
previous difficulties) and symplectic
groups on quaternions and complex linear quaternions. Appling a quaternionic
group theory to elementary particle physics, our purpose is to obtain a set
of groups for translating from complex to quaternionic quantum fields and to
emphasize the potentialities of the quaternionic groups for focusing on a
special class of GUTs. We conclude this brief introduction with an amusing
storical note. Quaternions were discovered by Hamilton~\cite{ham} on 16
October 1843. The Irish mathematician was so impressed by the new idea that
he scratched the main formula of the new algebra on a stone bridge that he
happened to be passing.

\section{Quaternionic algebra and complex geometry}

The quaternionic algebra over a field $\cal F$ is a set
\begin{equation} \label{h}
{\cal H} = \{\alpha + i\beta + j\gamma + k\delta \mid \alpha, \beta,
\gamma, \delta \in {\cal F}\}
\end{equation}
with operation of multiplication defined according to the
following rules for imaginary units:
\[i^{2} = j^{2} = k^{2} = -1\]
\[ij=k \; , jk=i \; , \;  ki=j \; ,\]
\[ji=-k \; , kj=-i \; , \;  ik=-j \; .\]
Complex numbers can be constructed from the real numbers by introducing a
quantity $i$ whose square is $-1$:
\[ c = r_{1} + i r_{2} \; \; \; \;
(r_{m} \in {\cal R} \; \; \; m=1,2) \; \; ;\]
likewise, we can construct the quaternions\footnote{Quaternions, as used in
this paper, will always mean {\em real quaternionic numbers} and never
{\em complexified quaternions} (${\cal F} = {\cal C}(1, {\cal I})$ in
eq.~(\ref{h}) with $\cal I$ which commutes with $i, \; j, \; k$). A number
of papers in the literature which
discuss quantum mechanics equations based on complexified quaternions can be
found in ref.~\cite{mor}.} from the complex numbers in exactly the same way
by introducing another quantity $j$ whose square is $-1$
\[ q = c_{1} + j c_{2} \; \; \; \;
(c_{m} \in {\cal C} \; \; \; m=1,2) \]
and anticommutes with $i$ ($ \{i, \; j\}=0 \Rightarrow k^{2}=-1$).\\
We need three imaginary units $i, \; j, \; k$ because
\[ ij=\alpha +i\beta +j\gamma \; \; \; , \; \; \; \alpha, \; \beta, \;
\gamma \in {\cal R} \; \; ,\]
implies
\[i^{2}j=i\alpha -\beta +ij\gamma=i\alpha -\beta +(\alpha +i\beta +j\gamma)
\gamma = ... + j \gamma^{2} \]
and so gives the inconsistent relation
\[\gamma^{2}=-1 \; \; .\]
In going from the complex numbers to the quaternions we lose the property of
commutativity. In going from the quaternions to the next more
complicated case (called octonionic numbers) we lose the property of
associativity. The situation can be graphically represented by the following
chart\cite{gil}

\begin{center}
\begin{tabular}{|c|c|c|c|c|c|} \hline
Name of     & Method of         & Real      & Division & Associa- & Commu-\\
field & construction & dimension & algebra     & tivity   & tativity\\ \hline
Real       & $r$                           & 1 & Yes & Yes & Yes \\ \hline
Complex    & $c=r_{1}+ir_{2}$              & 2 & Yes & Yes & Yes \\ \hline
Quaternionic & $q=c_{1}+jc_{2}$              & 4 & Yes & Yes & No \\ \hline
Octonionic   & $o=q_{1}+\tilde{\imath}q_{2}$ & 8 & Yes & No & No \\ \hline
\end{tabular}
\end{center}
We can immediatly shown the nonassociativity of the octonionic numbers in the
previous ``split'' representation. We have seven imaginary units
(the new imaginary unit $\tilde{\imath}$ anticommutes with the quaternionic
imaginary units $i, \; j, \; k$)
\[ i\; , \; j \; , \; k \; , \; \tilde{\imath}\; , \; I=i\tilde{\imath} \; , \;
J=j\tilde{\imath}\; , \; K=k\tilde{\imath} \; \; .\]
It is straightforward to verify
that
\[ IJk=JKi=KIj=-1 \]
\[ (IJk=i\tilde{\imath}j\tilde{\imath}k=
-i\tilde{\imath}^{\; 2}jk=ijk=k^{2}=-1) \; . \]
Associativity is dropped by following relations
\[i(\tilde{\imath}k)=-iK=J \; \; , \; \; (i\tilde{\imath})k=Ik=-J \; \; .\]

To complete this introduction to the quaternionic algebra, we introduce
the qua\-ternion conjugation operation denoted by $^{+}$ and defined by
\begin{equation}
1^{+}=1 \; \; \; \; \; (i, \; j , \; k)^{+} = - \; (i, \; j , \; k) \; \; .
\end{equation}
The previous definition implies
\[ (\psi \varphi)^{+} = \varphi^{+} \psi^{+} \; \; , \]
for $\psi$, $\varphi$ quaternioninc functions. The definition of a
conjugation operation which does not reverse the order of $\psi$, $\varphi$
factors is given by
\begin{equation} \label{*}
1^{*}=1 \; \; \; \; \; (i, \; j , \; k)^{*} = j^{+}(i, \; j , \; k)j \; \; .
\end{equation}

Remembering the noncommutativity of the quaternionic multiplication we must
specify whether the quaternionic Hilbert space $V_{\cal H}$ is to be formed
by right or by left multiplication of vectors by scalars. Besides we must
specify if our scalars are quaternionic, complex or real numbers. We will
follow the usual choice (see Adler~\cite{adl2}, Horwitz and
Biedenharn~\cite{hor}) to
work with a linear vector space under right multiplication by scalars.

Operators which act on states {\em only} from the left as in
\[ {\cal O} \mid \psi > \]
are named quaternion linear operators, they obey
\begin{equation}
{\cal O} (\mid \psi > q) = ({\cal O} \mid \psi >) q \; \; ,
\end{equation}
for an arbitrary quaternion $q$. More general classes of operators such as
complex or real linear operators can be introduced. We will use the notation
${\cal O}_{c}$ (${\cal O}_{r}$) to indicate complex (real) linear operators.
They act on quaternionic states as follows
\begin{equation}
{\cal O}_{c} (\mid \psi > c) = ({\cal O}_{1} + {\cal O}_{2} \mid i)
(\mid \psi > c) = ({\cal O}_{c} \mid \psi >) c \; \; ,
\end{equation}\begin{equation}
{\cal O}_{r} (\mid \psi > r) = ({\cal O}_{1} + {\cal O}_{2} \mid i +
{\cal O}_{3} \mid j + {\cal O}_{4} \mid k)(\mid \psi > r) =
({\cal O}_{r} \mid \psi >) r \; \; ,
\end{equation}
for an arbitray complex $c$, real $r$ (${\cal O}_{1, \; 2, \; 3, \; 4}$
represent quaternion linear operators).\\
The {\em barred} operators ${\cal O}\mid b$ act on quaternionic objects
$\psi$ as in
\[ ({\cal O}\mid b) \phi \: \equiv \: {\cal O} \phi b \; \: .\]

There are three scalar products which can be used to define a real-valued norm
$\parallel \psi \parallel$. We will call the binary mapping
$<\psi \mid \varphi>$ of $V_{\cal H} \times V_{\cal H}$ into $\cal H$, defined
by
\[ <\psi \mid \varphi> \; =\int d^{3}x \; \psi^{+} \varphi \; \; , \]
the quaternion scalar product (see Adler~\cite{adl}), and the binary mapping
$<\psi \mid \varphi>_{c}$ of $V_{\cal H} \times V_{\cal H}$ into $\cal C$,
defined by
\begin{equation}
<\psi \mid \varphi>_{c} \; =
\; \frac{1 - i\mid i}{2} \; <\psi \mid \varphi> \; \; ,
\end{equation}
the complex scalar product or complex geometry (as named by Rembieli\'nski in
ref.~\cite{rem}). The complex scalar product, used by  Horwitz
and Biedenharn~\cite{hor}, in order to define consistently multiparticle
quaternionic states, has then been applied in papers on the Dirac
equation~\cite{rot}, representations of $U(1,q)$~\cite{del1} and translations
between Quaternion and Complex Quantum Mechanics~\cite{del2}.\\
The last trivial possibility is represented by a real scalar product, the
binary mapping $<\psi \mid \varphi>_{r}$
of $V_{\cal H} \times V_{\cal H}$ into $\cal R$, defined by
\[ <\psi \mid \varphi>_{r} \; = \;
\frac{1 - i\mid i - j\mid j - k\mid k}{4} \; <\psi \mid \varphi> \; \; . \]
In this paper we will use a linear quaternionic Hilbert space under right
multiplication by complex scalars and will work with  complex scalar products.

To conclude
this section we recall that since $<\psi \mid \varphi>_{c}$ is  the complex
${\cal C}(1, \; i)$ projection of $<\psi \mid \varphi>$, any transformation
which is an invariance of $<\psi \mid \varphi>$ is automatically an invariance
of $<\psi \mid \varphi>_{c}$ as well. Obviously a transformation which is an
invariance of $<\psi \mid \varphi>_{c}$ is {\em not} automatically
an invariance of
$<\psi \mid \varphi>$. An example of that is given (see ref.~\cite{del3} or
for a brief review, section IV of this paper) by the quaternionic version of
the electroweak group $U(1, \; q) \mid U(1, \; c)$. This group represents an
invariance of $<\psi \mid \varphi>_{c}$ but not of $<\psi \mid \varphi>$.

\section{A new possibility}

In this section we give a {\em new} panoramic review of
quaternionic groups. Why {\em new}? As elements of our
matrices (given any two vector spaces $V_{n}$, $V_{m}$, every linear
operator $\cal O$ from $V_{n}$ to $V_{m}$ can be represented by an $m \times n$
matrix) we will not use simple quaternions but complex linear quaternions or
generalized quaternions as called in our previous works~\cite{del4}
\begin{equation} \label{a}
q_{c} = q_{1} + q_{2} \mid i  \; \; \; \; \; (q_{1, \; 2} \in {\cal H}) \; \; .
\end{equation}
Corresponding to our convention that $V_{\cal H}$ is a linear vector space
under right multiplication by complex scalars, the most general linear
one-dimensional operator which acts on quaternionic functions is in fact
represented by (\ref{a}).\\
The product of two complex linear quaternions $q_{c}$ and $p_{c}$, in
terms of quaternions $q_{1}, \; q_{2}, \; p_{1}$ and $p_{2}$, is given by
\[q_{c} p_{c} = q_{1}p_{1}-q_{2}p_{2} + (q_{1}p_{2} + q_{2}p_{1}) \mid i
\; \; . \]

Before discussing the groups $Gl(n, \; q_{c})$, we introduce a new
definition of transpose for quaternionic matrices which will allow us to
overcome previous difficulties (our definition which applies to standard
quaternions will be extended to complex linear quaternions).\\
The customary convention of defining the transpose $M^{t}$ of the matrix $M$ is
\[ (M^{t})_{rs}=M_{sr} \; \; .\]
In general, however, for quaternionic matrices $MN$ one has
\[ (MN)^{t} \neq N^{t} M^{t} \; \; ,\]
whereas this statement holds as an equality for complex matrices. Defining an
appropriate transpose for quaternionic numbers (which go back to usual
definition for complex number $c^{t}=c$) we can overcome the just cited
difficulty. The new transpose $q^{t}$ of the quaternionic number
\[ q = \alpha + i\beta + j\gamma + k\delta \; \; \; \; \; (\alpha, \beta,
\gamma, \delta \in {\cal R}) \]
is
\begin{equation}
q^{t} = \alpha + i\beta - j\gamma + k\delta \; \; .
\end{equation}
The transpose of a product of two quaternions $q$ and $p$ is the product of
the transpose quaternions in reverse order (note that $q^{t} = - j q^{+} j$)
\[ (qp)^{t} = p^{t} q^{t} \; \; . \]
Our convention of defining the transpose $M^{t}$ of the matrix $M$ is
\[ (M^{t})_{rs}=M_{sr}^{\; \; \; t} \]
and so we have
\[ (MN)^{t} = N^{t} M^{t} \; \; .\]
Remembering the $^{*}$ conjugation defined in (\ref{*}) we can
write
\[ M^{+}=M^{* \; t} \; \; .\]

Noting that under the transpose and quaternion conjugation operation
we have $i^{\; t}=i$ and $i^{\; +}=-i$, we can immediatly generalize
the definition of transpose and quaternion conjugation to complex
linear quaternions as follows
\[ q_{c}^{t} = q_{1}^{t} + q_{2}^{t} \mid i  \]
\[ (q_{c}^{+} = q_{1}^{+} - q_{2}^{+} \mid i) \; \; . \]
Introducing complex linear quaternions we create new possibilities in
quaternio\-nic quantum mechanics with complex geometry. For example we can
always trivially relate an anti-Hermitian operator $A$ to an Hermitian
operator $H$ by removing a factor $1 \mid i$
\begin{equation} \label{H}
H=A \mid i
\end{equation}
\[(<A\psi \mid \varphi>_{c} \; = - <\psi \mid A\varphi>_{c} \; \Rightarrow \]
\[-i<A\psi \mid \varphi>_{c} \; = i <\psi \mid A\varphi>_{c} \; = \;
<\psi \mid A\varphi>_{c} i \Rightarrow \]
\[<H\psi \mid \varphi>_{c}\; = \;<\psi \mid H\varphi>_{c}) \; \; . \]
This statement is not trivial in quaternionic quantum mechanics with
quaternionic geometry (see Adler ref.~\cite{adl} pag.~33).\\
In the literature we know an operator like that in (\ref{H}), the momentum
operator
\[ - \vec{\partial} \mid i \; \; , \]
given by Rotelli in his paper on the quaternionic Dirac equation~\cite{rot}.

The classical groups which occupy a central place in group representation
theory and have many applications in various branches of mathematics and
physics are the unitary, special unitary, orthogonal and symplectic groups.
So we will discuss in this paper the $U(n, \; q_{c})$, $SU(n, \; q_{c})$,
$O(n, \; q_{c})$, $Sp(n, \; q_{c})$ subgroups of $Gl(n, \; q_{c})$.

With complex linear quaternions we have the possibility to give a new
definition of trace by
\[ tr \; q_{c} = re(q_{1}) + i \; re(q_{2}) \]
which implies that for any two complex linear quaternions $q_{c}$ and $p_{c}$
\[ tr \; (q_{c}p_{c}) = tr \; (p_{c}q_{c}) \; \; . \]
We know that the generators of the unitary, special unitary, orthogonal and
symplectic groups must satisfy the following
conditions\footnote{A detailed classification of the real Lie algebras of
linear Lie groups is given by Cornwell in the book of ref.~\cite{cor}
(vol.~2, pag.~392).}
\begin{eqnarray*}
U(n) \; : & A + A^{+} =0 \; \; \; ,\\
SU(n)  \; : & A + A^{+} =0 \; \; , \; \; tr \; A =0 \; \; ,\\
O(n) \; : & A + A^{t} =0 \; \; ,\\
Sp(2n) \; : & {\cal J}A + A^{t}{\cal J} = 0 \; \; ,
\end{eqnarray*}
where
\[{\cal J}_{2n\times 2n} =
\left( \begin{array}{cc} 0_{n\times n} &  1_{n\times n}\\
$-1$_{n\times n} & 0_{n\times n}\end{array} \right) \; \; . \]
So for the generators of the one-dimensional groups with complex linear
quaternions we have
\begin{center}
\begin{tabular}{rcccl}
$U(1, q_{c})$     & : & $q_{c}+q_{c}^{+}=0$ & $\Rightarrow$ &
$A=i, \; j, \; k, \; 1 \mid i  \; \; \; ;$  \\
$SU(1, \; q_{c})$ & : & $tr \; q_{c} = 0 $ & $\Rightarrow$ &
$A=i, \; j, \; k \; \; \; \; ; $ \\
$O(1, \; q_{c})$  & : & $ q_{c}+q_{c}^{t}=0$  & $\Rightarrow$ &
$A=j, \; j \mid i  \; \; \; \; ;$\\
$Sp(1, \; q_{c})$  & : & $ jq_{c}+q_{c \;}^{t \;}j=0$  & $\Rightarrow$ &
$ A = i, \; j, \; k, \; i \mid i, \; j \mid i, \; k \mid i \; \; .$
\end{tabular}
\end{center}
Any complex linear quaternion group of dimension $n$ is isomorphic to a
complex representation of dimension $2n$. We give the transformation rule
(for further detail see ref.~\cite{del2})
\[ \left( \begin{array}{cc} c_{1} & c_{2}\\ c_{3} & c_{4} \end{array} \right)
\left( \begin{array}{c} z_{1}\\ z_{2} \end{array} \right) \Longleftrightarrow
\left[ \frac{c_{1} + c_{4}^{*}}{2} + j \frac{c_{3} - c_{2}^{*}}{2} +
(\frac{c_{1} - c_{4}^{*}}{2i} + j\frac{c_{3} + c_{2}^{*}}{2i}) \mid i \right]
(z_{1}+jz_{2}) \; \; . \]
Remembering that a complex linear quaternions, in terms of real quantities,
is expressed by
\[q_{c} = \alpha_{1} +i\beta_{1} +j\gamma_{1} + k\delta_{1} +
(\alpha_{2} +i\beta_{2} +j\gamma_{2} + k\delta_{2}) \mid i \]
\[ \alpha_{1, \; 2}, \; \beta_{1, \; 2}, \; \gamma_{1, \; 2}, \;
\delta_{1, \; 2} \in {\cal R} \; \; ,\]
we have
\[ complex \; linear \; quaternions \supset quaternions \supset
complex \; \; , \]
and more
\[ complex \; linear \; quaternions \supset elements \; like \; \alpha_{1} +
\alpha_{2} \mid i \equiv c_{right} \; \; .\]
So
\[ Gl(n \; q_{c}) \supset Gl(n \; q) \supset Gl(n \; c) \; \; , \]
\[ Gl(n \; q_{c}) \supset Gl(n \; c_{right}) \; \; . \]

We can now give the general formulas for counting the generators of generical
$n$-dimensional groups as a function of $n$.
\begin{center}
$\diamond$ Dimensionalities of groups $\diamond$
\end{center}
\begin{center}
\begin{tabular}{|lcccc|} \hline
$U(n, \; q_{c})$      & : & $4n+8 \; \frac{n(n-1)}{2}$ & = & $4n^{2}$  \\
$U(n, \; q)$          & : & $3n+4 \; \frac{n(n-1)}{2}$ & = & $n(2n+1)$ \\
$U(n, \; c_{right})$  & : &  $n+2 \; \frac{n(n-1)}{2}$ & = & $n^{2}$ \\ \hline
\end{tabular}
\end{center}
\begin{center}
\begin{tabular}{|lcc|} \hline
$SU(n, \; q_{c})$  & : & $4n^{2}-1$ \\
$SU(n, \; q)$  & $\equiv$ & $U(n, \; q)$\\
$SU(n, \; c_{right})$  & : & $n^{2}-1$ \\ \hline
\end{tabular}
\end{center}
\begin{center}
\begin{tabular}{|lcccr|} \hline
$O(n, \; q_{c})$      & : & $2n+8\; \frac{n(n-1)}{2}$  & = & $2n(2n-1)$ \\
$O(n, \; q)$          & : &  $n+4 \; \frac{n(n-1)}{2}$  & = & $n(2n-1)$  \\
$O(n, \; c_{right})$  & : &  $ 2 \; \frac{n(n-1)}{2}$  & = & $ n(n-1)$\\ \hline
\end{tabular}
\end{center}
For the quaternionic symplectic groups we have
\[{\cal J}_{2n\times 2n} =
\left( \begin{array}{cc} 0_{n\times n} &  1_{n\times n}\\
$-1$_{n\times n} & 0_{n\times n}\end{array} \right) \; \; \; , \; \; \;
 {\cal J}_{(2n+1)\times (2n+1)} =
\left( \begin{array}{ccc} 0_{n\times n} & 0_{n\times 1} &  1_{n\times n}\\
0_{1\times n} & j & 0_{1\times n}\\
$-1$_{n\times n} & 0_{n\times 1} & 0_{n\times n}\end{array} \right) \; \; ,\]
so
\begin{center}
\begin{tabular}{lcccc}
$Sp(2n, \; q_{c})$      & : &
$8n^{2}+2 \; [ \; 6n+8 \; \frac{n(n-1)}{2} \; ]$  & = & $4n(4n+1)$ \\
$Sp(2n, \; q)$          & : &
$4n^{2}+2 \; [ \; 3n+4 \;\frac{n(n-1)}{2} \; ]$  & = & $2n(4n+1)$ \\
$Sp(2n, \; c_{right})$  & : &
$2n^{2}+2 \; [ \; 2n+2 \; \frac{n(n-1)}{2} \; ]$  & = & $2n(2n+1)$ \\
$Sp(2n+1, \; q_{c})$      & : &
$4n(4n+1)+2(8n)+6$  & = & $2(2n+1) \; [ \; 2(2n+1)+1 \; ]$ \\
$Sp(2n+1, \; q)$      & : &
$2n(4n+1)+2(4n)+3$  & = & $(2n+1) \; [ \; 2(2n+1)+1 \; ]$
\end{tabular}
\end{center}
The situation for the symplectic groups can be summarize as follows
\begin{center}
\begin{tabular}{|lcc|} \hline
$Sp(n, \; q_{c})$  &     :    & $2n(2n+1)$ \\
$Sp(n, \; q)$      & $\equiv$ & $U(n, \; q)$ \\
$Sp(2n, \; c_{right})$  & : & $2n(2n + 1)$ \\ \hline
\end{tabular}
\end{center}

\section{Quaternionic groups for GUTs}

Finally we can apply quaternionic group theory to elementary
particle physics. We must remark on an important point. A symmetry operation
$\cal S$ of a system, described by $\mid \psi>$, is a mapping of $\mid \psi>$
into $\mid \psi'>$, which preserves all transition probabilities
\[ {\cal S}\mid \psi> = \mid \psi'> \]
\[ \mid <\varphi'\mid \psi'> \mid^{2} =
\mid <\varphi\mid \psi> \mid^{2} \; \; . \]
In quaternionic quantum mechanics with complex geometry
a `{\em quaternionic} $\mid$ {\em compl\-ex}' phase
\begin{equation} \label{t}
e^{i\alpha + j\beta + k\gamma} \mid e^{i\delta}
\end{equation}
appears. We can immediatly prove that the previous transformation represents
an invariance of $<\psi \mid \varphi>_{c}$
\[ <\psi' \mid \varphi'>_{c} = e^{-i\delta} <\psi \mid \varphi>_{c}
e^{i\delta} = <\psi \mid \varphi>_{c} \]
(the transformation (\ref{t}) obviously does not represent an invariance of
the quaternionic scalar product $<\psi \mid \varphi>$).
So a quaternionic invariance group
like that of the electroweak gauge group (for further details on the group
$U(1, \; q)$ in quaternionic quantum mechanics with complex geometry see
ref.~\cite{del1}) naturally appears.
In a recent work~\cite{del3} we have studied the Higgs
sector of the electroweak model from the point of view of quaternionic
quantum mechanics with complex geometry. The Higgs fields are assumed to be
four (two complex) and this coincides with the number of solutions of the
standard Klein-Gordon equation within quaternionic quantum mechanics with
complex geometry. The global invariance group of the one-component
Klein-Gordon equation is $U(1, \; q) \mid U(1, \; c)$ isomorphic at the Lie
algebra level with the Glashow-Salam-Weinberg group.

The aim of this paper is to extend our previous consideration about
quaternionic electroweak models and to propose quaternionic groups for GUTs.

Within our formalism the peculiarity is the
doubling of solutions (note that with complex scalar products
$\mid \psi >$ and $\mid \psi >j$ are orthogonal states), so we have some
problems in discussing the color group (three states: $R$, $G$, $B$).\\
There are three possibilities. The first one represents a {\em conservative}
hypothesis, the second and the third ones represent interesting ideas with
potential {\em predictive} powers.
\begin{itemize}
\item $SU(3, \; c_{right})$ for color group.\\
We have the following doubling of states
\[ \left( \begin{array}{c} R\\ G\\ B \end{array} \right) \; \; \; , \; \; \;
j\left( \begin{array}{c} R\\ G\\ B \end{array} \right) \; \; \; \; \; . \]
We need a `new' quantum number to differentiate the previous solutions.
The appropriate quantum number is represented by the weak isospin. So we
can rewrite the previous solutions as follows
\[ \left( \begin{array}{c} u_{R}\\ u_{G}\\ u_{B} \end{array} \right) \; \;
\; , \; \; \; j\left( \begin{array}{c} d_{R}\\ d_{G}\\ d_{B}
\end{array} \right) \; \; \; \; \; . \]
Note that the {\em complex} group $SU(3, \; c_{right})$ does not mix $u$ with
$jd$, besides the
{\em one-dimensional} quaternionic group $U(1, \; q)$ does not mix $R$, $G$,
$B$. We are particularly pleased with that. So the color group
$SU(3, \; c_{right})$ suggests the weak-isospin group $U(1, \; q)$.
The gauge group for the standard model is\footnote{In our Lagrangian we
need a complex
projection~\cite{del3}  for a Dirac Langrangian density in order to obtain
the Dirac field equation and so the complex group $U(1, \; c_{right})$
will be always an invariance of our Lagrangian.}
\[ SU(3, \; c_{right}) \times U(1, \; q)_{L} \times U(1, c_{right})_{Y} \]
(in this way using the color group $SU(3, \; c_{right})$ we have a
translation between complex and quaternionic theories).
\item $SU(3, \; c)$ for color group.\\
We have always a doubling of states, but in this case the complex solutions
transform like $3$ whereas the $j$-complex solutions like $3^{*}$ (to see
that it is sufficient to note that $ij=ji^{*}$). So working with the
standard group $SU(3, \; c)$ we remark the possibilty of additional
multiplets.\\ The minimal grand unification group $SU(5, \; c)$~\cite{geo} will
have (in our formalism) the following additional multiplets
\[ 5 \; \; + \; \; j5^{*} \; \; , \]
\[ 10 \; \; + \; \; j10^{*} \; \; . \]
This is an interesting result,
in fact we know that a single unification point cannot be
obtained within minimal (non-supersymmetric) $SU(5, \; c)$. The
$\alpha_{strong}$ coupling misses the crossing point of the other two by
more than eight standard deviations. In the `quaternionic' version of
$SU(5, \; c)$ additional multiplets of quark and leptons naturally appear and
so we could find right unification proprietis. In a work of Amaldi et
al.~\cite{ama} a non-supersymmetric $SU(5, \; c)$ model, based on
additional split multiplets (split multiplet models also
appear in ref.~\cite{fra}), is proposed. Their model shows unification
properties similar to the minimal supersymmetric extension of the standard
model
\begin{center}
\begin{tabular}{lclr}
$M_{thresold}$ & = & $10^{3.2  \pm 0.9}$ & $Gev \; \; ,$\\
$M_{GUT}$      & = & $10^{16.0 \pm 0.3}$ & $Gev \; \; .$
\end{tabular}
\end{center}
\item Quaternions for color group.\\
Looking at charts of the previous section we can immediatly observe that the
minimal quaternionic group candidate for color group is
\[SU(2, \; q_{c}) \; \; .\]
In fact its $15$ generators contain the $8$ generators of the standard
color group $SU(3, \; c)$. In this case we have not a doubling of solutions,
nevertheless we must note the appearance of an additional solution
\[ \left( \begin{array}{c} R+jG\\ B+jW \end{array} \right) \; \; . \]
In this case we start with  the gauge group $SU(2, \; q_{c})$ and break down
to the usual color group (in the quaternionic version). We need a {\em fourth}
color.
\end{itemize}
What about the fourth color? The idea of a fourth color (as lepton number) was
proposed in $1973$ by Pati and Salam~\cite{pat}. With quarks and leptons in
one multiplet of a local gauge symmetry group $G$, baryon and lepton number
conservations cannot be absolute. This line of reasoning had led Pati and
Salam to predict in their paper~\cite{pat} that the lightest baryon - the
proton - must ultimately decay into leptons. The $PS$ model was proposed
before any grand unification scheme and so it constitutes really the forerunner
of the GUT idea that quarks and leptons should belong to common
representations of the gauge group.\\
Following the $PS$ idea we can put the fermions of the first
generation in the following multiplets
\[ \left( \begin{array}{c} u_{R}+j u_{G}\\ u_{B}+j \nu_{W} \end{array} \right)
\; \; \; , \; \; \;
\left( \begin{array}{c} d_{R}+jd_{G}\\ d_{B}+je_{W}
\end{array} \right) \]
and propose an `electrostrong' model based on the gauge group
\[ SU(2, \; q_{c}) \times U(1, \; c_{right}) \; \; .\]

If we wish to consider unification in the context of a bigger
gauge group than $SU(5 \; c)$ we must consider the group $SO(10, \; r)$. This
group can break down to the standard model gauge group in many different
chains of symmetry breaking. Chains preferred from the CERN LEP data include
the $PS$ model (for further details see Deshpande et al.~\cite{des},
Galli~\cite{gal}). We can now immediatly translate~\cite{del2} the $PS$
model based on the complex group
\[ SU(4) \times SU(2)_{L} \times SU(2)_{R} \]
by the quaternionic group
\begin{equation} \label{ps}
SU(2, \; q_{c}) \times U(1, \; q)_{L} \times U(1, \; q)_{R}
\end{equation}
and propose a GUT model based on the group $O(5, \; q)$ which represents the
minimal quaternionic group which contains the gauge group (\ref{ps}).

We conclude this paper with a completly new idea {\em inspired} by
quaternions. We have discussed the problem concerning the odd number of
colors. Before $1974$, discovery of $J/\psi$ (bound state of a charmed
quark and a charmed antiquark) at Brookhven National Laboratory and at
Stanford Linear Accelerator Center, we would have had the same problem with
the flavor group. In that case the correct predictive
hypothesis\footnote{The conservative hypothesis is represented by using
$SU(3, \; c_{right})$ for the flavor group, with the spin as new
quantum number to differentiate the doubling of solutions
\[ \left( \begin{array}{c} u_{\uparrow}\\ d_{\uparrow}\\ s_{\uparrow}
\end{array} \right) \; \;
\; , \; \; \; j\left( \begin{array}{c} u_{\downarrow}\\ d_{\downarrow}\\
s_{\downarrow} \end{array} \right) \; \; \; \; \; .\]} should have been
the choice of $SU(2, \; q_{c})$ for flavor group and the choice of a new quark
as fourth flavor
\[ \left( \begin{array}{c} u+j d\\ s+j c \end{array} \right)
\; \; . \]
So why do we not propose a white quark as fourth color? This possibility is
currently under investigation~\cite{pap}. An interesting quaternionic group
for GUTs (only
proposed here) appears natural if we believe in the existence of white quarks.
The just cited quaternionic groups is
\[ SU(3, \; q_{c}) \; \; .\]
This group represents the natural quaternionic extension of the
group $SU(5, \; c)$, is algebrically isomorphic to $SU(6, \; c)$
and contains the (new) color group and the electroweak group. In our
forthcoming paper we will focus our attention on this group. Remembering
that the anomaly for the representation $R$ may be characterized by
\[ tr \; (\{T^{a}, \; T^{b} \}T^{c})=A(R)  \; d^{abc} \; \; , \]
\[ (T^{a} \; normalized \; generators \; of \; the \;
representations \; R \; \; ,\]
\[d^{abc} \; symmetric \; structure \; constants \;  of \;  the \;  Lie \;
algebra) \; ,\]
and that for a representation given by the completely antisymmetric product
of $p$ fundamental representations of $SU(n, \; c)$, the coefficient $A(R)$ is
\[ A(R) = \frac{(N-3)!(N-2p)}{(N-p-1)!(p-1)!} \; \; , \]
we have for $SU(6, \; c)$ (the complex counterpart of $SU(3, \; q_{c})$) an
anomaly cancellation when
\[ p = \frac{N}{2} = 3 \; \; . \]
That implies three vertical boxes in the Young tableaux and so a
$20$ dimensional
representation. We now have $20$ particles to accomodate in this
representation, in fact we can add to the standard $16$ particles of the
first generation the following {\em new} four particles
\[ u_{W \; L}, \; u_{W \; L}^{c}, \; d_{W \; L}, \; d_{W \; L}^{c}  \; \; .\]

A last possibility concerning quaternion groups for GUTs is given
by the choice of
the quaternionic group $SU(3, \; q_{c})$, but without requiring a fourth color.
The unification of the standard coupling constants could appear through the
split-multiplet mechanism for the complementary heavy fermions. The complex
counterpart of $SU(3, \; q_{c})$, namely $SU(6, \; c)$, is considered in detail
in an interesting work of Chkareuli et al.~\cite{chk}. We briefly summarize
their results:
\begin{center}
$SU(6, \; c)$ model with
\end{center}
\begin{itemize}
\item one family of complementary fermions
\[ M_{intermediate \; breaking}=5.4\times 10^{2} \; Gev \; \; \; ,
\; \; \; M_{GU}=1.3\times 10^{16} \; Gev \; \; \; , \]
\item two families of complementary fermions
\[ M_{intermediate \; breaking}=2.4\times 10^{9} \; Gev \; \; \; ,
\; \; \; M_{GU}=1.2\times 10^{16} \; Gev \; \; \; , \]
\item three families of complementary fermions
\[ M_{intermediate \; breaking}=3.9\times 10^{11} \; Gev \; \; \; ,
\; \; \; M_{GU}=1.3\times 10^{16} \; Gev \; \; \; . \]
\end{itemize}

\section{Conclusions}

In this paper we have given an informal panoramic review of the
quaternionic groups. Our aim was to analyse possible quaternionic groups for
GUTs. We have obtained a set of groups for translating from standard complex
quantum fields to a particular version of quaternionic quantum fields
and have proposed some new groups with potential predictive powers. In the
following charts we list our results.

\begin{center}
$\diamond$ Groups for translating from {\em cqm} to {\em qqm} with complex
geometry $\diamond$
\end{center}
\begin{center}
\begin{tabular}{|cc|} \hline
$SU(3, \; c_{right}) \times U(1, \; q)_{L} \times U(1, \; c_{right})$  &
standard model\\ \hline
$SU(2, \; q_{right}) \times U(1, \; q)_{L} \times U(1, \; q)_{R}$  &
$PS$ standard model\\ \hline
$O(5, \; q)$  &
$SO(10, \; r)$ GUT model\\ \hline
$SU(3, \; q_{c})$  &
split-multiplets provide \\
  & for unification\\ \hline
\end{tabular}
\end{center}

\begin{center}
$\diamond$ Groups with potential predictive power $\diamond$
\end{center}
\begin{center}
\begin{tabular}{|cc|} \hline
$SU(5, \; c)$  &
split-multiplets,\\
  & proposed by Amaldi et al. ,\\
  & naturally appear\\ \hline
$SU(3, \; q_{c})$  &
flavor {\em inspires} fourth color,\\
  & hypothetical existence of\\
  & white quarks\\ \hline
\end{tabular}
\end{center}

We have proposed some quaternionic groups for GUTs and remark on
their potentialities for focusing on a special class of standard complex
GUTs (a detailed review of the complex groups for unified model
building is given in ref.~\cite{lan}). A further analysis of the
quaternionic groups introduced here will be given in a more detailed work,
where we will particularly focus our attention on the quaternionic
group $SU(3, \; q_{c})$.

Finally we wish to remember that we have another possibility to look at
fundamental physics as proposed by Harari-Shupe~\cite{har}. We can think of
quarks and leptons as composites of other more fundamental fermions,
{\em preons}. A stimulating idea (within quaternionic
quantum mechanics with quaternionic geometry) about this possibility
is proposed by Adler (see~\cite{adl3} or~\cite{adl} pag.~501). He suggests
that the color degree of freedom postulated in the Harari-Shupe scheme could
be sought in a noncommutative extension of standard quantum mechanics.

We hope that this paper emphasizes the nontriviality in the
choice to adopt quaternions as the underlying number field and remarks on the
possibile predictive power in using new mathematical formalism to describe
theoretical physics\footnote{``The most powerful method of advance that can
be suggested at present is to employ all the resources of pure mathematics in
attemps to perfect and generalize the mathematical formalism that forms the
existing basis of theoretical physics, and after each succes in this
direction, to try to interpret the new mathematical features in terms of
physical entities'' - Dirac~\cite{dir}.}. Why $i$, $j$ , $k$~? Why not~?

\section*{Acknowledgments}
The the very nice hospitality of the Institute for Advanced Study,
where this paper was prepared, is gratefully acknowledged. The author is
deeply indebted to Steve Adler for his important suggestions and for
stimulating conversations, and thanks Pietro Rotelli for a correspondence
containing useful notes.

\end{document}